\documentclass[sn-mathphys,Numbered]{sn-jnl}
\usepackage{graphicx}
\usepackage{multirow}
\usepackage{amsmath,amssymb,amsfonts}
\usepackage{amsthm}
\usepackage{mathrsfs}
\usepackage{xcolor}
\usepackage{textcomp}
\usepackage{manyfoot}
\usepackage{booktabs}
\usepackage{algorithm}
\usepackage{algorithmicx}
\usepackage{algpseudocode}
\usepackage{listings}
\usepackage{subcaption}
\usepackage{upgreek}
\usepackage{adjustbox}
\usepackage{multirow}
\usepackage{graphicx,bbold}
\usepackage{titletoc}

\titlecontents{chapter}             
  [0em]                             
  {\vspace*{\baselineskip}}         
  {\bfseries\thecontentslabel\quad} 
  {}                                
  {\quad\thecontentspage}           
  []                                
\titlecontents*{section} 
  [0em]                  
  {\space}               
  {\thecontentslabel~}   
  {}                     
  {}                     
  [{\mdseries\hspace{0.4em}\titlerule*[6pt]{.}}\contentspage\\]              
\titlecontents*{subsection} 
  [0em]                     
  {\space}                  
  {}                        
  {}                        
  {}                        
  [.\space]                 

\newcommand{\trm}[1]{\textrm{#1}}
\newcommand{\tbf}[1]{\textbf{#1}}
\newcommand{\tsf}[1]{\textsf{#1}}

\newcommand{\figref}[1]{Fig.\,\ref{#1}}
\newcommand{\figrefa}[1]{Fig.\,\ref{#1}a}
\newcommand{\figrefb}[1]{Fig.\,\ref{#1}b}
\newcommand{\secref}[1]{Sec. \ref{#1}}
\newcommand{\bea}{\begin{eqnarray}}
\newcommand{\eea}{\end{eqnarray}}
\newcommand{\bi}{\begin{itemize}}
\newcommand{\ei}{\end{itemize}}
\newcommand{\alphaqed}{\alpha}
\newcommand{\Eqed}{F_{\tsf{qed}}}

\def\lambdabar{\protect\@lambdabar}
\def\@lambdabar{%
\relax \bgroup
\def\@tempa{\hbox{\raise.73\ht0
\hbox to0pt{\kern.2\wd0\vrule width.7\wd0
height.1pt depth.1pt\hss}\box0}}%
\mathchoice{\setbox0\hbox{$\displaystyle\lambda$}\@tempa}%
{\setbox0\hbox{$\textstyle\lambda$}\@tempa}%
{\setbox0\hbox{$\scriptstyle\lambda$}\@tempa}%
{\setbox0\hbox{$\scriptscriptstyle\lambda$}\@tempa}%
\egroup }

\definecolor{bkcol}{rgb}{0.14,0.42,0.9}

\begin{document}

\title{Input to the European Strategy for Particle Physics: Strong-Field Quantum Electrodynamics\vspace{-0.5\baselineskip}}

\author[1]{G.~Sarri}
\author[2]{B.~King}
\author[3]{T.~Blackburn}
\author[4]{A.~Ilderton}

\author[5]{S.~Boogert}
\author[6]{S.S.~Bulanov}
\author[7]{S.V.~Bulanov}

\author[8]{A.~Di Piazza}
\author[9]{L.~Ji}
\author[10]{F.~Karbstein}
\author[11]{C.\,H.~Keitel}

\author[12]{K.~Krajewska}

\author[13]{V.~Malka}
\author[14]{S.P.D.~Mangles}

\author[15]{F.~Mathieu}
\author[16]{P.~McKenna}
\author[17]{S.~Meuren}
\author[18]{M.~Mirzaie}
\author[19]{C.~Ridgers}
\author[20]{D.~Seipt}
\author[21]{A.G.R.~Thomas}
\author[22]{U.~Uggerh\o j}
\author[23]{M.~Vranic}
\author[24]{M.~Wing}

\affil[1]{School of Mathematics and Physics, Queen's University of Belfast, BT7 1NN, Belfast, UK}
\affil[2]{Centre for Mathematical Sciences, University of Plymouth, Plymouth, PL4 8AA, UK}
\affil[3]{Department of Physics, University of Gothenburg, 41296, Gothenburg, Sweden}
\affil[4]{Higgs Centre, School of Physics and Astronomy, University of Edinburgh, Edinburgh EH9 3FD, UK}

\affil[5]{Cockcroft Institute, Daresbury Laboratory, STFC, Keckwick Lane, Daresbury, WA4 4AD, Warrington, UK}

\affil[6]{Lawrence Berkeley National Laboratory, Berkeley, CA 94720, USA}

\affil[7]{ELI Beamlines Facility, The Extreme Light Infrastructure ERIC, Doln\'i B\u re\u zany, Czech Republic}

\affil[8]{Department of Physics and Astronomy, University of Rochester, Rochester, New York 14627, USA; Laboratory for Laser Energetics, University of Rochester, Rochester, New York 14623, USA}

\affil[9]{State Key Laboratory of High Field Laser Physics, Shanghai Institute of Optics and Fine Mechanics (SIOM), Chinese Academy of Sciences (CAS), Shanghai 201800, China}

\affil[10]{Theoretisch-Physikalisches Institut, Abbe-Center of Photonics, Universität
Jena, 07743 Jena, Germany}

\affil[11]{Max Planck Institute for Nuclear Physics, Saupfercheckweg 1, D 69117 Heidelberg, Germany}
\affil[12]{Institute of Theoretical Physics, Faculty of Physics, University of Warsaw, 02-093 Warsaw, Poland}

\affil[13]{Department of Physics of Complex Systems, Weizmann Institute of Science, Rehovot 7610001, Israel}
\affil[14]{The John Adams Institute for Accelerator Science, Imperial College London, London, SW7 2AZ, UK}

\affil[15]{Laboratoire pour l’Utilisation des Lasers Intenses, CNRS, Ecole Polytechnique, Palaiseau, France}
\affil[16]{Department of Physics, SUPA, University of Strathclyde, Glasgow G4 0NG, UK}
\affil[17]{SLAC National Accelerator Laboratory, Menlo Park, CA 94025, USA}
\affil[18]{Center for Relativistic Laser Science, Institute for Basic Science, Gwangju 61005, Korea}
\affil[19]{York Plasma Institute, School of Physics, Engineering and Technology, University of York, Heslington, UK}
\affil[20]{Helmholtz Institute Jena, Fr\"obelstieg 3, 07743 Jena, Germany}
\affil[21]{Center for Ultrafast Optical Science, University of Michigan, Ann Arbor, MI 48109-2099, USA}
\affil[22]{Department of Physics and Astronomy, Aarhus University, 8000 Aarhus, Denmark}
\affil[23]{Golp/Instituto de Plasma e Fus\~ao Nuclear, Instituto Superior T\'ecnico, Universidade de Lisboa, 1049-001 Lisbon, Portugal}
\affil[24]{Department of Physics and Astronomy, University College London, London, UK}

\abstract{This document sets out the intention of the strong-field QED community to carry out, both experimentally and numerically, \textbf{high-statistics parametric studies of quantum electrodynamics in the non-perturbative regime}, at fields approaching and exceeding the critical or `Schwinger' field of QED ($\Eqed = m^2c^3/e\hbar \approx 1.3 \times 10^{18}$ V/m). In this regime, several exotic and fascinating phenomena are predicted to occur that have never been directly observed in the laboratory. These include Breit-Wheeler pair production, vacuum birefringence, and quantum radiation reaction. This experimental program will also serve as a stepping stone towards studies of elusive phenomena such as elastic scattering of real photons and the conjectured perturbative breakdown of QED at extreme fields. State-of-the-art high-power laser facilities in Europe and beyond are starting to offer unique opportunities to study this uncharted regime at the intensity frontier, which is highly relevant also for the design of future multi-TeV lepton colliders. However, a transition from qualitative observational experiments to quantitative and high-statistics measurements can only be performed with large-scale collaborations and with systematic experimental programs devoted to the optimisation of several aspects of these complex experiments, including detector developments, stability and tolerances studies, and laser technology.}

\maketitle

\thispagestyle{empty}

\clearpage 

\setcounter{page}{1}

\rmfamily
\section{Scientific Context} \label{sec:Intro}
\begin{figure}[h!!]
\centering
\includegraphics[width=6cm]{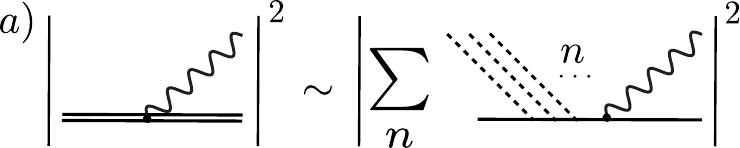}\hspace{1.5cm} \includegraphics[width=8.5cm]{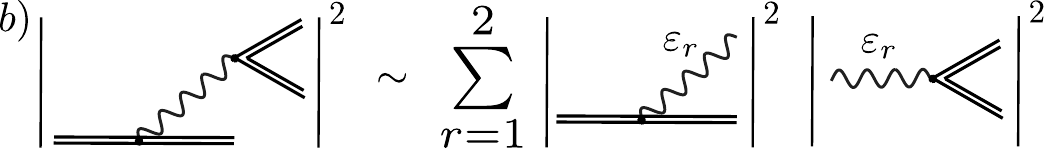}
\vspace{0.2cm}
\caption{a): Fermion states are `dressed' in the EM background, which can be expressed as a sum over all numbers of exchanges of Fourier momenta with the background (dashed lines). b): in the `long-pulse limit' higher-order dressed processes are well-approximated by sequences of first-order dressed processes where the intermediate particle is on-shell and polarised.} \label{fig:NLC}
\end{figure}

\vspace{-2mm}
Quantum Electrodynamics (QED) is a part of the Standard Model that has been tested with experiments to a precision of better than one part in a trillion \cite{Hanneke:2010au, Aoyama:2017uqe,Fan2023}. Such precision tests of QED are in a regime where the coupling $\alphaqed$ is small and the accuracy of theory calculations can be increased by adding successive orders of radiative corrections. In this regime, the electromagnetic (EM) field is incoherent. The situation can be quite different though when studying QED in coherent EM fields, where the coherence can enhance the  fundamental coupling $\alphaqed$. A well-understood example is in atomic physics, where the effective coupling in hydrogenic ions of charge number $Z$ becomes $Z\alphaqed$. When $Z$ is large, it leads to a test of nonperturbative QED in strong atomic fields, most lately with a precision of better than one part in a billion \cite{Morgner2023}. For spacetime-varying coherent fields such as slowly-varying magnetic fields \cite{Kaspi:2017fwg}, laser pulses \cite{Di_Piazza_2012,Gonoskov:2021hwf,Fedotov:2022ely}, those in beam-beam collisions \cite{DelGaudio:2018lfm,Yakimenko:2018kih,Taya:2024wrm}, beam-laser collisions as well as internal fields around symmetric structures in oriented single crystals \cite{Uggerhoj_2005,Wistisen:2017pgr}, the effective coupling can be written $\alphaqed \xi^{2n}$, where $\xi$ is the so-called \emph{intensity parameter} and $n$ is the number of interactions with the coherent background. If the EM background is a single laser pulse, which is only weakly focussed, its interaction with electrons and positrons can be well-described as interactions with a classical plane wave background \cite{Fedotov:2022ely}. For a plane wave of field strength $F$, the intensity parameter can be written as (assuming $\epsilon_0=\hbar=c=1$):
\[
\xi = \frac{eF}{m\omega} \equiv \frac{\trm{work done by field on charge over a Compton wavelength}}{\trm{energy of field quantum}}.
\]
Hence $\xi$ gives a measure of the typical number of equivalent `photons' (the background is classical) interacting with a charge, $e$, over its Compton wavelength $\lambdabar = 1/m$ (we take $m$ to be the electron mass). When $\xi\ll1$, the effect of the background field is \emph{perturbative} and the probability is well-approximated by including just the minimum number of charge-field interactions for the process at hand to occur. For example, the total probability of Compton scattering when $\xi\ll1$ is well-approximated by including just a single electron-background interaction. For electrons and positrons, values of $\xi > 1$ can now routinely be probed in experiments \cite{Sarri:2014gea,Yan:2017,Mirzaie:2024iey,Cole:2017zca,Poder:2017dpw,Los:2024} and all orders of interaction between the charge and the field must be included (see e.g. \figrefa{fig:NLC}). For electrons and positrons colliding with a laser pulse, the squared intensity parameter can be written as \cite{Abramowicz:2019gvx} $\xi^{2} \approx 2\alphaqed \lambdabar^{2}\lambda_{l}n_{l}$ where $\lambda_{l}$ is the laser wavelength and $n_{l}$ is the density of equivalent laser photons. Hence $\xi \gtrsim 1$ is an example of \tbf{non-perturbativity at small coupling} similar to gluon saturation in deep inelastic scattering and the $x\to 0$ limit of the Bjorken factor \cite{bjorken1969asymptotic} in QCD.
\vspace{-0.5cm}

To perform calculations at arbitrary $\xi$ parameter, the Dirac equation is solved in a given EM background $F_{\tsf{bg}}$ and scattering processes are calculated using these solutions and the quantised part of the EM field $f$. `Volkov' fermion wavefunctions corresponding to a plane wave EM are routinely employed in calculations. In the `Furry picture' \cite{Furry_1951}, the quantum field $f$ is a perturbation but the interaction with the background $F_{\tsf{bg}}$ is included to all orders. 

Unlike in `standard' QED with incoherent fields, the tree-level three-point vertex \emph{is} in general kinematically allowed in a coherent plane-wave background. The example of nonlinear Compton scattering $e^{\pm}\to e^{\pm}+\gamma$ is shown in \figrefa{fig:NLC}. The coherent field breaks the translational symmetry of the vacuum so that four-momentum conservation no longer holds. For example, in a plane wave EM field, only the momenta transverse and momentum parallel to wavevector $k^{\mu}$ (`lightfront momentum') are conserved. Therefore it is useful to define for a given particle of momentum $p^{\mu}$ and mass $m$ its \emph{energy parameter}: 
\[
\eta = \frac{k\cdot p}{m^2} \equiv \trm{frequency of the background in the rest frame of a particle of mass $m$ and momentum $p$}.
\] 
In the case of nonlinear Compton scattering, the energy parameter of the incoming electron is then equal to the sum of energy parameters of the emitted electron and photon. Another example of a three-point vertex is nonlinear Breit-Wheeler pair-creation $\gamma \to e^{+}e^{-}$ (depicted in the right-most diagram of \figrefb{fig:NLC}), which is a purely quantum effect requiring the \emph{strong-field parameter} $\chi$ to satisfy $\chi \gtrsim 1$. In a plane wave and for electrons and positrons, the strong-field parameter $\chi$ is equal to the field strength in the rest frame of the charge, $F_{\tsf{rf}}$ in units of the QED field scale, $\Eqed=m^{2}/e$ (also known as the `Schwinger Limit')  i.e. $\chi = F_{\tsf{rf}}/\Eqed$. \tbf{Strong-field QED} (SFQED) is then accessed when $\chi \gtrsim 1$.  The strong-field parameter can also be written in terms of the work done as:
\[
\chi = \frac{eF_{\tsf{rf}}}{m^2} \equiv \frac{\trm{work done by field on charge in its rest frame over a Compton wavelength}}{\trm{rest mass of particle}}
\]
and hence when $\chi \gtrsim 1$, pair-creation via nonlinear Breit-Wheeler can be accessed experimentally. In a plane-wave background, the strong-field parameter is the product of intensity and energy parameters, i.e. $\chi=\xi \eta$. Since $\chi \propto \hbar$ whereas $\xi$ is classical, $\chi$ can be used to quantify the magnitude of nonlinear quantum effects. 

In some regions of parameter space, strong-field QED processes can display a non-perturbative dependence on $\chi$. For example, in the quasi-static regime when $\xi\gg1$ and $\chi \ll 1$, the probability for nonlinear Breit-Wheeler pair-creation scales as $\sim \exp(-8/3\chi)$. If $\chi$ is written in terms of the QED field strength scale $\Eqed$ explicitly, this scaling becomes $\sim \exp[-(8m/3\eta k^0)(\Eqed/F)]$, which has a similar dependency on the field strength $F$ as Schwinger pair-creation $\sim \exp[-\pi\, \Eqed/F]$. Since $\Eqed \sim 1/ \sqrt{\alpha}$, this limit can be referred to as \tbf{non-analytic pair creation}, which displays a non-perturbative dependency on the coupling. Thus a further comparison with QCD can be made: string-breaking in the process of meson creation can be modelled using the Schwinger mechanism \cite{casher1979chromoelectric}, which displays a similar behaviour to nonlinear Breit-Wheeler in the quasi-static limit.

Lightfront momentum conservation has a consequence for QED processes of a higher order in dressed vertices, such as the second-order nonlinear trident process $e^{\pm}\to e^{\pm}+e^{+}e^{-}$ depicted in \figrefb{fig:NLC}. If the spacetime overlap of the probe with the coherent background is long enough, the probability is dominated by the contribution from the photon propagator going on-shell such that the second-order process can be well-approximated by sequential first-order processes (in this case, nonlinear Compton scattering of a photon followed by the photon's transformation into an electron-positron pair via nonlinear Breit-Wheeler pair-creation). In the so-called `long-pulse limit', probabilities are dominated by processes with on-shell propagators. This disrupts the usual perturbative hierarchy in $\alpha$ and reduces the relative contribution of radiative corrections. 
\begin{figure}[h!!]
\centering
\includegraphics[width=9cm]{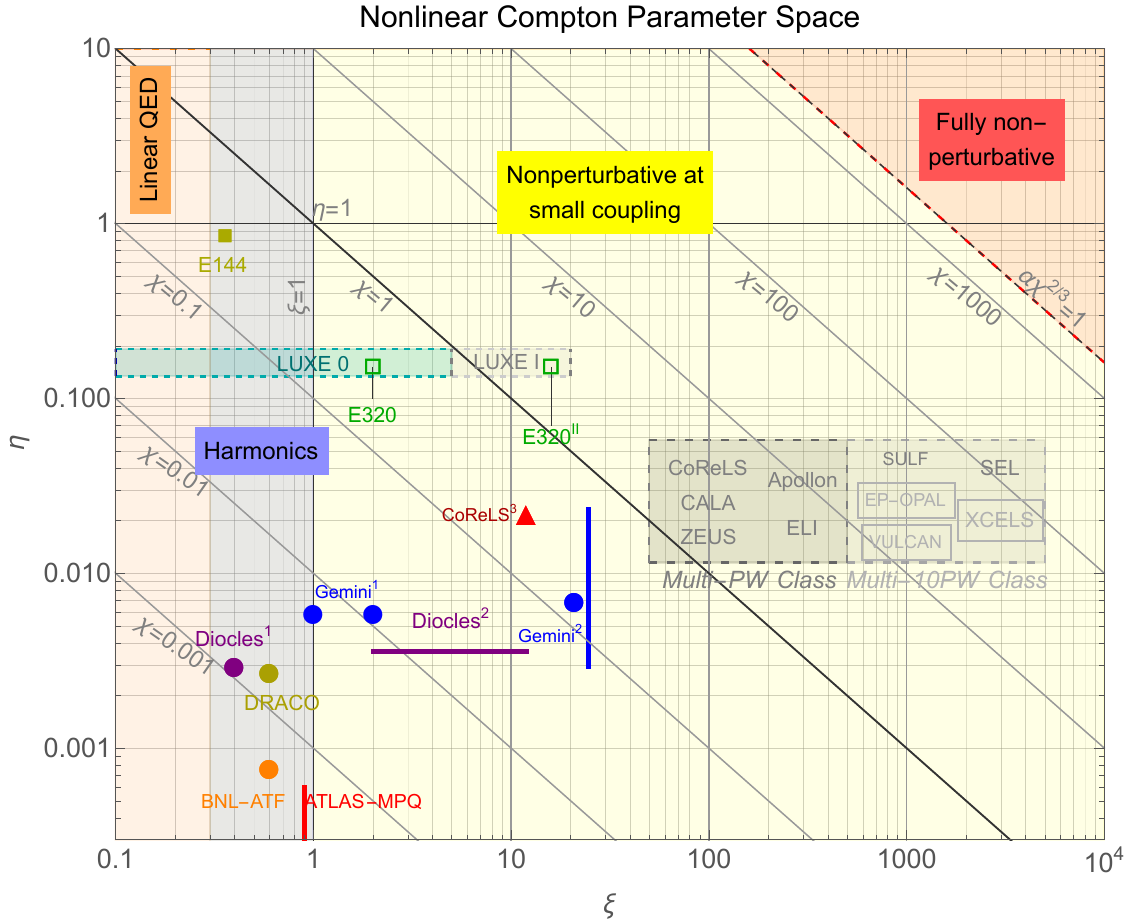}\includegraphics[width=9cm]{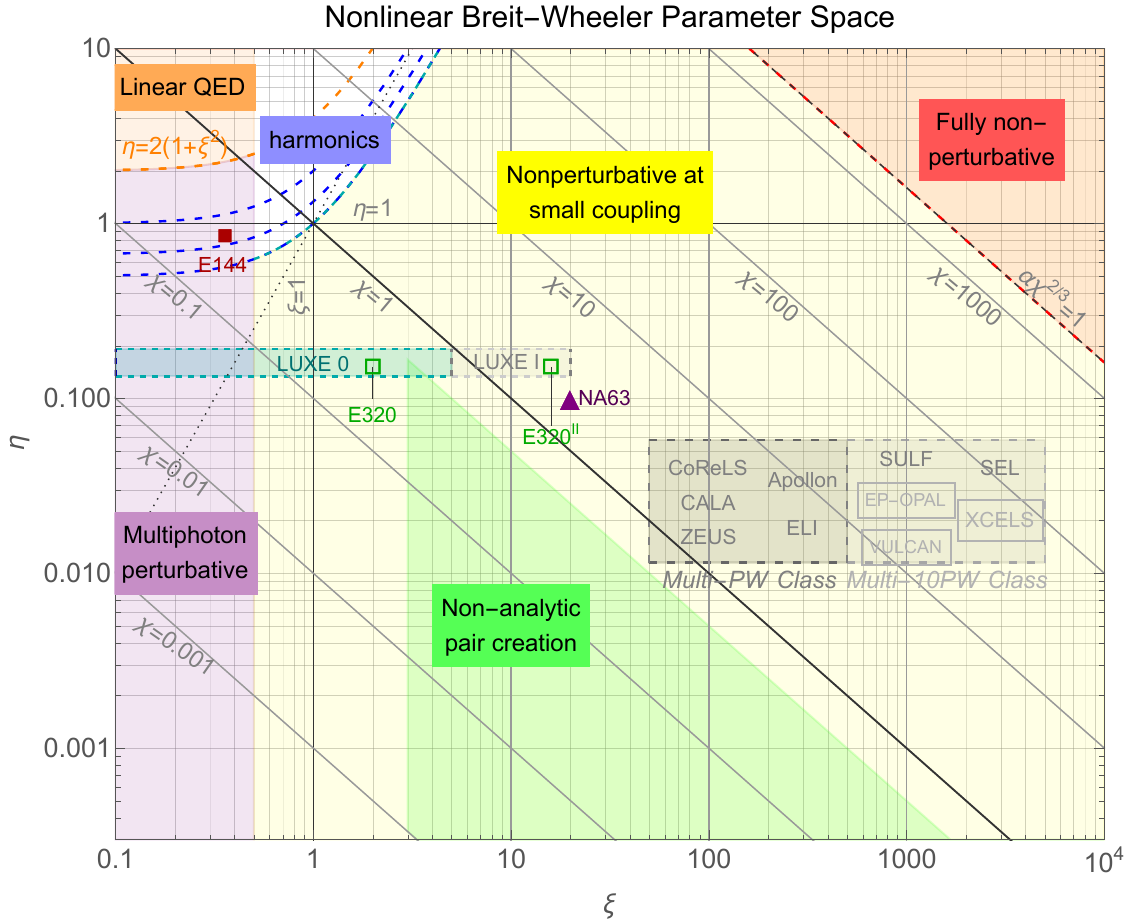}\vspace{0.2cm}
\caption{Nonlinear Compton (left) and nonlinear Breit-Wheeler (right) parameter spaces (adapted from \cite{Fedotov:2022ely}). Solid lines and markers indicate reported experimental results; dashed lines and empty markers indicate planned experiments. The grey `Multi-PW Class' and `Multi-10PW Class' regions correspond to typical values of $\xi$ that can be produced at these laser facilities, and values of $\eta$ correspond to $1\ldots5\,\trm{GeV}$ particle beams colliding at 20 degrees with the laser pulse, na\"ively assuming the electron reaches the peak intensity at the focus of the laser pulse. The specific experimental values plotted, in $(\xi,\eta)$ co-ordinates, are as follows. \textbf{Left:} 
      Apollon \cite{papadopoulos2016}; ATLAS-MPQ $(0.9,5\ldots 6\times 10^{-4})$ \cite{2015PhRvL.114s5003K}; BNL-ATF $(0.6, 7.5\times10^{-4})$ \cite{Sakai:2015mra}; CALA \cite{2021NJPh...23j5002S}; CoReLS \cite{sung20174}; DIOCLES$^1$ $(0.4,0.0029)$ \cite{Chen:2013mba}; DIOCLES$^2$ $(2\ldots12,0.0036)$ \cite{Yan:2017}; DRACO $(0.6,0.0027~[230\,\trm{MeV}])$  \cite{Hannasch:2021kyh}; E144 $(0.36,0.83)$ \cite{Bamber:1999zt}; E320 $(2,0.15)$ \cite{Salgado:2021fgt}; E320$^{\tsf{II}}$ (proposed upgrade) $(16,0.15)$ \cite{meuren2019probing}; ELI \cite{Turcu:2016dxm}; {EP-OPAL \cite{Zuegel:14}}; Gemini$^{1}$ $(1\ldots2,0.006)$ \cite{Sarri:2014gea}; Gemini$^{2}$ $(24.7,0.003\ldots0.02)$ \cite{Cole:2017zca,Poder:2017dpw}; LUXE 0 $(0.1\ldots5,0.13\ldots0.19)$ \cite{Abramowicz:2021zja,LUXE:2023crk}; LUXE 1 $(5\ldots20,0.10\ldots0.19)$ \cite{Abramowicz:2021zja,LUXE:2023crk}; {VULCAN \cite{V2020}}; SEL \cite{sel18}; {SULF \cite{Li:18}}; XCELS \cite{Mukhin_2021}; ZEUS \cite{Nees21}. 
    \textbf{Right:} E144 $(0.36,0.83)$ \cite{Bamber:1999zt}, NA63 $(20,0.1)$ \cite{Nielsen:2022bws}. (Both experiments measured nonlinear trident in a regime well-approximated as having nonlinear Breit-Wheeler pair-creation as the second of two steps.)
}\label{FIG:PARAMPLOTS}
\end{figure}

As the intensity and duration of the EM background is increased, it can become more probable for an incoming charge or photon to scatter multiple times in a \emph{cascade} of processes. In a \emph{shower} cascade, the background has a negligible effect on the charges' trajectory and the background acts in a similar manner to a dense target producing bremsstrahlung. In \emph{avalanche} cascades~\cite{Mironov:2014xba} the background has a significant effect on charges' trajectory, e.g. by reaccelerating charges that lost energy to emission. An exponentially rising number of particles can also change the local plasma density and lead to background field depletion \cite{Grismayer2016}. If the colliding particle is an electron, the classical limit is known as (classical) \emph{radiation reaction}. 
Classical radiation reaction is formally described by the Lorentz-Abraham-Dirac equation (for reviews see~\cite{Burton03042014,Blackburn:2019rfv}),
the solutions of which suffer from pre-acceleration and runaways, signalling the breakdown of classical theory over very short distance scales. Its better-behaved reduction-of-order approximation, the Landau Lifshitz equation~\cite{landau.lifshitz}, is believed to be a good approximation for all experimentally-accessible scenarios~\cite{DiPiazza:2018luu,Ekman:2021eqc,Nielsen:2020ufz,Nielsen:2021ppf}. Deviations from the classical limit are described by \emph{quantum radiation reaction} and signatures of these deviations are searched for in experiment \cite{Cole:2017zca,Poder:2017dpw,Los:2024}. 
At very high values of the strong-field parameter satisfying $\alpha\chi^{2/3}\sim O(1)$, the Furry picture is predicted to break down \cite{Fedotov:2016afw}. According to the Ritus-Narozhny conjecture, in EM backgrounds that can be regarded as locally constant and crossed (transverse), the contribution to the probability of a certain class of (dressed) loop diagrams scales as $\alpha\chi^{2/3}$, and therefore when this becomes of order unity, the perturbative expansion in the quantum photon field breaks down and QED becomes \tbf{fully non-perturbative}. The parameter space 
for nonlinear Compton scattering and nonlinear Breit-Wheeler are depicted in 
\figref{FIG:PARAMPLOTS}.

\vspace{-2mm}

\section{International landscape and motivation}\label{Sec2}
\vspace{-2mm}
\begin{table*}[h!]
\centering
\begin{tabular}{|l|c|c|c|c|c|}
\hline
\textbf{FACILITY} & \textbf{COUNTRY} & \textbf{E (J)} & \textbf{$\tau$ (fs)} & \textbf{$\xi$} & \textbf{Rep. rate (Hz)}\\
 \hline
 FACET-II \cite{meuren2019probing} & USA & 0.3 & 30 & 6 & 10\\
 GEMINI \cite{GEMINI} & UK & 15 & 45 & 32 & 1/20\\
 BELLA \cite{BELLA} & USA & 40 & 40 & 55 & 1\\
 EPAC \cite{EPAC} & UK & 30 & 30 & 55 & 10\\
 CALA \cite{CALA} & DE & 70 & 27 & 90 & 1\\
 ZEUS \cite{Mukhin_2021} & USA & 75 & 25 & 100 & 1/60\\
 APOLLON \cite{papadopoulos2016} & FR & 70 & 22 & 100 & 1/60 \\
 CoReLS \cite{sung20174} & KR & 80 & 20 & 110 & 1/10 \\
  ELI-NP \cite{ELINP} & RO & 200 & 20 & 180 & 1/60\\
SULF \cite{SULF} & CN & 300 & 23 & 200 & 1/60\\  
 ELI-BL \cite{ELIBL} & CZ & 1500 & 150 & 230 & 1/60\\
 \hline
 \hline
\emph{NSF OPAL} \cite{OPAL} & USA & 2$\times$500 & 20 & 280 & 1/300\\
\emph{VULCAN2020} \cite{V2020} & UK & 400 & 20 & 330 & 1/300\\
\emph{SEL} \cite{Yujie:2021}  & CN & 25,000 & 25 & 1,800 & / \\
 \hline
\end{tabular}
\caption{Summary of the main high-power laser facilities in operation or in design (in italics) capable of hosting SFQED experiments. For all facilities the value for $\xi$ is indicative, estimated assuming a focal spot with FWHM = 2 $\mu$m containing 60\% of the laser energy.
}
\label{table1}
\end{table*}
\vspace{-0.5cm}
With the advent of chirped pulsed amplification \cite{Strickland:2018} (that led to the award of the Nobel prize in physics in 2018), the achievable peak power of laser pulses has been experiencing a fast-paced increase. Several PW-class laser systems are now operational worldwide, with a handful of laser facilities operating at the 10 PW now entering the international landscape \cite{Danson:2019} (Table 1). If tightly focussed, these laser systems can generate peak intensities in excess of $10^{23}$ Wcm$^{-2}$ ($\xi > 100$ \cite{2021Optic...8..630Y}). These lasers can now be coupled with GeV-scale electron beams from either radio-frequency accelerators or from plasma-based wakefield accelerators. In principle, this experimental capability enables accessing quantum parameters in excess of unity: for example, a 5 GeV electron beam interacting head-on with a focussed laser beam with a peak intensity of $10^{23}$ Wcm$^{-2}$ would result in a quantum parameter $\chi\approx 10$. 
The high degree of tuneability of the laser and electron beam parameters can also allow for extensive parametric studies across several regimes of SFQED. 

Stemming from proof-of-principle experiments carried out at SLAC in the mid-90s \cite{Burke:1997ew,Bula:1996}, pilot experiments in the area have recently reported novel and intriguing hints of strong-field phenomena. Studies of nonlinear Compton scattering \cite{Sarri:2014gea,Yan:2017,Mirzaie:2024iey}, quantum radiation reaction \cite{Cole:2017zca,Poder:2017dpw,Los:2024}, and Breit-Wheeler pair production \cite{Kettle:2021} have been recently reported, demonstrating the possibility of accessing SFQED in the laboratory. It is now necessary to progress to the next stage of scientific investigation, stepping from pilot experiments to systematic and high-statistics studies. 
This is necessary to benchmark and refine existing analytical and numerical models, it will significantly contribute to advancing our understanding of fundamental science in a scarcely explored regime, and will provide disruptive advancements in several areas of basic and applied sciences. SFQED phenomena are thought to play a dominant role at the interaction point of the next generation of particle colliders \cite{Yakimenko:2018kih} and in the dynamics and radiative properties of astrophysical plasmas in proximity of ultra-massive objects \cite{Kaspi:2017fwg}. SFQED phenomena will also play a significant role in the dynamics of laser-driven plasmas, thus strongly affecting the properties of novel particle and radiation sources (see, e.g., Ref. \cite{Tamburini:2010}). Several high-power laser facilities (see, e.g., Table \ref{table1}) have studies of SFQED as a central topic in their scientific agenda.  Some examples of phenomenology that can be experimentally accessed are given below:

\textbf{2A) Nonlinear Compton scattering} involves a charged  particle emitting a photon during interaction with the laser background. It can be written schematically as:
\begin{equation}
e^- + n\gamma_l \rightarrow e^- + \gamma
\end{equation}
where $n$ is the harmonic number (net number of equivalent laser photons). The magnitude of nonlinear and quantum effects in this process can be directly observed in the spectrum of the scattered electrons and the emitted photons, in particular by measuring the position of the `Compton edge' (end of the kinematic range of the first harmonic), which is sensitive to the laser intensity $\xi$ and the strong-field parameter $\chi$ and must thus be identified with a high spectral resolution (of the order of, or less than, a percent).

\textbf{2B) Nonlinear Breit-Wheeler pair production}
is the inelastic counterpart to the scattering of real photons, where a photon is transformed into an electron-positron pair. It can be written as:
\begin{equation}
\gamma + n\gamma_{l} \rightarrow e^- + e^+ 
\end{equation}
where $n$ is the laser harmonic order. This process is sometimes referred to as \emph{linear} Breit-Wheeler (BW) when the centre-of-momentum energy is high enough that the threshold for creating an electron-positron pair is already satisfied with $n=1$. Linear BW pair production was recently observed in Coulomb fields in ultra-peripheral heavy ion collisions involving quasi-real photons \cite{STAR:2019wlg}, but has never been directly observed during the collision of real photons. Linear BW can be triggered during the interaction of GeV-scale photons (e.g., from bremsstrahlung or inverse Compton scattering of an ultra-relativistic electron) with keV-scale x-ray photons (as, for instance, produced during the the interaction of a high-energy laser beam with a thin solid target), as first attempted in \cite{Kettle:2021} and numerically studied in different configurations (see, e.g., Refs. \cite{Pike:2014,Ribeyre2016}). Nonlinear BW can be triggered e.g. during the interaction of bremsstrahlung or Compton photons with the field of a focussed intense laser pulse, which, as mentioned in \secref{sec:Intro}, can lead to an experimental signal of non-analytic pair-creation \cite{Hartin:2018sha,King:2024ffy}.

\textbf{2C) Nonlinear trident pair production} is the leading-order tree-level production of an electron-positron pair by an electron $e^- \to e^{-}+e^{-}e^{+}$, see Fig.~\ref{fig:NLC} (b). Because a laser pulse is extended in spacetime compared to e.g. the EM fields in a heavy ion collision (which are relativistically contracted), the contribution from the photon propagator being on-shell is expected to dominate for most realistic experimental scenarios. This means that nonlinear trident can be approximated as a combination of the previous two processes: nonlinear Compton scattering of a real photon followed by its propagation and subsequent nonlinear BW pair production. In terms of laser harmonics:
\begin{equation}
e^- + n\gamma_l \rightarrow e^- + \gamma \hspace{5mm} \trm{and} \hspace{5mm} \gamma + n'\gamma_l \rightarrow e^{-}e^+ 
\end{equation}
This process whas been observed in the perturbative regime ($\xi\ll1$) \cite{Burke:1997ew}. The nonperturbative regime ($\xi\gg 1$) has been explored theoretically \cite{Hu2010,Dinu:2017uoj,Dinu:2019wdw} and
observed during the collision of 200 GeV electrons
with oriented crystals \cite{Nielsen:2022bws}, but questions still remain about the contribution when the photon propagator is off-shell \cite{Ilderton:2010wr,Nielsen:2023icv}.

\textbf{2D) Radiation reaction} broadly refers to the cumulative recoil experienced by an accelerated charged particle when it emits photons~\cite{Blackburn:2019rfv}. Classically, radiation reaction is consistently accounted for by the Landau-Lifshitz equation~\cite{landau.lifshitz}. In this case, the energy loss of an electron with Lorentz factor $\gamma$ can be expressed as~\cite{thomas.prx.2012}:
\begin{equation}
\frac{\Delta\gamma}{\gamma} = \frac{\sqrt{\frac{\pi}{2}}\tau_0\tau_l\omega_l^2\gamma\xi^2}{1+\sqrt{\frac{\pi}{2}}\tau_0\tau_l\omega_l^2\gamma\xi^2},
\end{equation}
where $\tau_0 = 2e^2/3m\simeq6.4\times10^{-24}$ s and $\tau_l$ is the laser pulse duration. 
When the field in the rest frame approaches the Schwinger field, two main quantum phenomena should be incorporated. First, in the classical limit there is no constraint on the frequency of the radiation emitted. However, in a quantum picture, the energy of the emitted photon cannot exceed that of the electron. This results in a reduction of the energy loss, which can be quantified by the so-called Gaunt factor. This factor can be expressed as the ratio between the spectral intensity of the quantum and classical synchrotron radiation. Useful approximations to the Gaunt factor are (\cite{thomas.prx.2012} and \cite{baier98}, respectively):
\begin{equation}
g(\chi) \approx (1+ 12 \chi +31 \chi^2 + 3.7\chi^3)^{-4/9}; \qquad g(\chi) \approx (1+4.8 (1+\chi)\ln (1+1.7 \chi) + 2.44 \chi^2 )^{-2/3}
\end{equation}
The latter has been compared to experiments colliding $O(100\,\trm{GeV})$ electrons with single crystals \cite{CERNNA63:2012zsc}).
Additionally, while emission of radiation is classically a continuous process, the emission of photons is probabilistic.
This implies that an electron can propagate further in the laser electromagnetic field without emitting (\emph{straggling} and \emph{quenching})~\cite{Blackburn:2014cig,Harvey:2016uiy}. Emission of gamma photons in this regime can further lead to spin-dependent deflection of electrons \cite{Li:2018fcz} and even twist of the electron wave-packet \cite{Bu24}. A strong observable is a broadening of the energy spectrum of the electrons after the interaction~\cite{Neitz:2013qba,Vranic:2015sft,niel.pre.2018}.
Quantum radiation reaction is extremely high order in dressed vertices, hence its theoretical interest. 
Laser-based experiments have provided strong initial evidence of this phenomenon~\cite{Cole:2017zca,Poder:2017dpw,Los:2024}, but precision measurements and validation of theoretical models and approximations are still outstanding.

\textbf{2E) QED cascades:} for sufficiently intense electromagnetic fields, an avalanche of electron-positron pair production can take place. The cascade can be sustained by nonlinear BW pair production and inverse Compton scattering, with a multiplicity that is directly linked to the laser intensity.  
This particular class of QED phenomena can result in the creation of dense pair plasma, a common occurrence in extreme astrophysical environments, such as neutron stars, pulsars, and magnetars \cite{Kaspi:2017fwg}. Several schemes have been proposed: irradiation of a solid target with an ultraintense light pulse \cite{Ridgers:2012}, collision of several laser pulses in vacuum \cite{Gelfer:2015ora,Vranic2017,PhysRevX.7.041003}, and collision of a laser pulse with a beam of relativistic electrons or high-energy photons \cite{Cole:2017zca,Poder:2017dpw,Los:2024,Kettle:2021}. A key experimental observable would be the number of electron-positron pairs generated, which is predicted to increase exponentially in time as the cascade forms.

\textbf{2F) Photon-photon scattering} is the elastic counterpart of the Breit-Wheeler process. Four-photon scattering is the elastic counterpart to the \emph{linear} BW process and as such is \emph{not} a SFQED process, but is the most discussed in the literature. When one or more outgoing photons are provided by an intense laser pulse, the strongest signal is from \emph{stimulated} photon-photon scattering, which can be written in terms of laser harmonics $n$ and $n'$ as:
\[
\gamma + n\gamma_{l} \to \gamma' + n'\gamma_{l}
\]
(with $n=n'=1$ for the four-photon process). Stimulated photon-photon scattering is enhanced by the coherence of the laser field \cite{Heinzl:2024cia} and has a different kinematics to the scattering of free photons. The emitted photon can be distinguishable from the incoming photon by having a modified momentum, energy and/or polarisation \cite{Lundstrom:2005za,DiPiazza:2006pr,Dinu:2013gaa,Karbstein:2015xra}. The cross-section of elastic photon-photon scattering with real photons is only loosely constrained experimentally \cite{Watt:2025} (although the ATLAS collaboration reported the observation of light-by-light scattering with quasi-real photons during ultra-peripheral heavy ion collisions \cite{ATLAS:2017}). 
An enhanced feasibility of detection has been put forward in a strongly asymmetric set-up of the two scattering photon sources in \cite{Sangal2021}. \emph{Nonlinear} photon-photon scattering is the elastic counterpart to nonlinear BW and \emph{is} a SFQED process but proposals for discovery currently require measuring the polarisation of GeV photons \cite{King:2016jnl,Macleod:2023asi,lv2024a}.

\textbf{2G) Vacuum polarisation:} if the strong-field parameter $\chi \ll 1$ the scattering of photons in an intense laser background can be calculated in effective field theory using the Heisenberg-Euler Lagrangian. The resulting modified photon dispersion relation can be described by adding a vacuum current and charge density to Maxwell's equations, so that the intense laser background appears to magnetise and polarise regions of the vacuum the photon propagates through (see e.g. \cite{King:2015tba,Fedotov:2022ely} for reviews). At the microscopic level, this corresponds to:
\[
\gamma_{1} + \gamma_{2} + \cdots \gamma_{j} \to \gamma_{1}' + \gamma_{2}' + \cdots \gamma'_{j'}.
\]
If written in a form that includes laser harmonics, $n\gamma_{l}$ would be added to one side and $n'\gamma_{l}$ to the other side of this equation.
The simplest case is leading-order photon-photon scattering $j=j'=1$, or `vacuum birefringence' \cite{Baier:1967,Heinzl:2006xc}. 
One way to measure this effect is to use a bright probe beam (such as an x-ray free electron laser (XFEL)) and enhance the process by colliding the probe x-ray photons with an  intense coherent EM field such as provided by a high power optical laser. This is one of the aims of the HIBEF collaboration \cite{HIBEF,Ahmadiniaz:2024xob}, which aims to detect such effects at the Eu.XFEL, and the SEL (Station of Extreme Light) in the SHINE project \cite{Shen:2018lbq}.
In order to make the polarisation flip in the x-ray photons  and hence the signal of vacuum birefringence detectable, it has been suggested to achieve very high polarisation purities \cite{Ahmadiniaz:2024xob}, develop x-ray polarimeters with extremely high precision \cite{Bernhardt:2020}, implement a sufficiently good `shadow factor' \cite{PhysRevLett.129.061802}, or use two optical beams to spatially separate the signal photons in the detector plane \cite{Ahmadiniaz:2022nrv,Macleod:2024jxl}. Beyond $j=j'=1$, interesting effects can occur at higher orders: $j>j'$ corresponds to photon merging or higher-harmonic generation; $j<j'$ corresponds to photon splitting, such as hypothesised around strongly-magnetised neutron stars \cite{Turolla:2015mwa}. 

Several alternative schemes (i.e., not laser-based) to experimentally study strong-field phenomena in this area have been reported. For example, a possible alternative is to study the dynamics of ultra-relativistic particles, and their radiative properties, as they propagate through aligned crystals, in the so-called channeling regime. Signatures of quantum effects in radiation reaction in this configuration have been recently reported~\cite{Wistisen:2017pgr}. 
\vspace{-2mm}
\section{Objectives of the SFQED input}
\vspace{-2mm}

Stemming from the proof-of-principle results mentioned above, it is necessary to \textbf{transition from qualitative observational experiments to quantitative and high-statistics precision measurements}. This demands a step-change improvement in experimental capability, numerical modelling, and theoretical understanding, which requires expertise across several areas of physics (e.g., lasers, accelerators, detector technology, plasma physics, and QED) in large-scale collaborations with programmatic funding. The main objectives are summarised below.
\vspace{-2mm}
\addcontentsline{toc}{section}{3.1}
\subsection{Experimental Objectives}
 \vspace{-2mm}
 As our overarching goal, we propose the \textbf{installation and operation of experimental platforms at high-power laser facilities, open to the growing user community in Europe and beyond, focussed on detailed and high-statistics study of SFQED phenomena}. Such an endeavour requires systematic studies on a wide range of complementary aspects of cutting-edge experimental and theoretical physics and investment in beamlines for high-power laser and electron beams. This work will also enable key technical developments with significant impact in other applications envisaged for high-power laser facilities, such as the generation of novel radiation and particle sources and proof-of-principle studies towards TeV-scale colliders. Key developments in experimental capability can be grouped as:

\textbf{3.1.A) Design and installation of dedicated beamlines and user areas at high-power laser facilities:} several facilities in Europe are equipped with high-power lasers capable of accessing SFQED regimes; however, dedicated investment is urgently required to ensure their effective use. Examples of this include, but are not limited to, beam transport to ensure co-location of different laser beams within the same experimental area, installation of high-power lasers at high-energy accelerators, construction of dedicated end-stations equipped with suitable diagnostics and detectors, and schemes for long-term programmatic access that extend beyond the typical duration of beam-time allocation in user facilities. 

\textbf{3.1.B) Control and diagnosis of high-intensity laser beams:} the electromagnetic fields of a tightly focussed high-intensity laser have complex spatiotemporal distributions with steep transverse and longitudinal gradients that significantly influence the dynamics and radiative properties of particle beams traversing it. Novel diagnostics for the detailed measurement of the spatiotemporal distribution of electromagnetic fields in the focus of a high-power laser have been recently proposed \cite{Pariente:2016}; however, it is necessary to ensure precise on-shot characterisation and control of these quantities. Ideally, one would need to achieve \%-level precision and shot-to-shot reproducibility on the field intensity in focus. This is necessary also for laser-solid interactions \cite{Ridgers:2012} and for the implementation of intensity-enhancing techniques such as relativistic mirrors (see, e.g., Ref. \cite{Zaim:2024}). 

\textbf{3.1.C) Control and diagnosis of high-energy electron beams:} for an effective interaction with the laser, electron beams must contain sufficient charge (10s of pC up to nC) in femtosecond-scale durations and micron-scale transverse sizes, while having well-characterised and controllable spectrum and chirp \cite{Magnusson:2023gbs}. Due to the invasive nature of the interaction, the electron beam properties prior to interaction are not readily accessible. While neural network techniques have been developed to circumvent this issue \cite{Streeter:2023} in laser-wakefield accelerators, a superior level of stability and control over the electron beams must be achieved, with pilot experiments showing potential in this direction \cite{Maier:2020}. While the coupling of a high-power laser with electron beams from a radio-frequency accelerator is expected to ease this issue, this type of experimental configuration is still not widely available, with only two experimental platforms worldwide currently operating or proposed \cite{meuren2019probing,LUXE:2023crk}.

\textbf{3.1.D) Active stabilisation of the interaction region over sustained periods of operation:} maintaining micron-scale spatial overlap and femtosecond-scale synchronisation between multiple high-power lasers or between a high-power laser and a high-energy particle beam is a phenomenally challenging task. However, it is a necessary ingredient to ensure sustained periods of operation and, thus, high statistics. Even a slight spatiotemporal misalignment can dramatically change the interaction conditions (see, e.g., Ref. \cite{Los:2024}). Active feedback loops and optics stabilisation techniques are starting to be implemented; however, detailed experimental programs devoted to this task must be carried out to ensure the high level of statistics required. Developments in this area will also be of particular benefit for the generation of novel radiation and particle sources and proof-of-principle studies towards TeV-scale colliders.

\textbf{3.1.E) Development of high-precision and high-sensitivity particle and photon detectors:} detectors with high-resolution operating in a high-background environment must be developed and thoroughly characterised. The detectors must be able to operate in a wide range of particle and photon energies and over a wide range of fluxes. Typical examples include single-particle detectors to characterise BW pair production \cite{Salgado:2021fgt}, ultra-high quality x-ray polarimeters to study vacuum birefringence \cite{Bernhardt:2020}, and high-flux gamma-ray detectors \cite{Fleck:2020,Cavanagh:2023,Fleck:2024,AVONI:2024}. These developments require large-scale collaborations with interdisciplinary expertise (see, e.g., the HIBEF and LUXE collaborations \cite{Ahmadiniaz:2024xob,LUXE:2023crk}).

\textbf{3.1.F) Novel strategies for background reduction and noise rejection:} several aspects of SFQED are associated with relatively rare events, thus producing weak signals in a high background. Advancing pilot work in this area (e.g., Ref. \cite{Salgado:2021fgt}), specific and sophisticated strategies for background reduction and for noise rejection at the detectors must thus be developed to ensure statistically meaningful measurements.

\textbf{3.1.G) New concepts to increase field intensity with existing laser facilities:}
new concepts to further enhance the field strength at the interaction point must be developed and experimentally tested. Recently proposed techniques include back-reflecting the laser onto the wakefield-accelerated electron beam \cite{TaPhuoc2012,Matheron2024}, and exploiting intensity boosts from relativistic plasma mirrors \cite{Fedeli:2021}. In parallel, current work on increasing the repetition rate of high-power laser facilities must be supported.

\addcontentsline{toc}{section}{3.2}
\subsection{Theory Objectives} \label{sec:theoryObj}
\vspace{-2mm}

\hspace{0.5cm}\textbf{ 3.2 A) Higher-multiplicity (dressed) processes:} in comparison to scattering in vacuum, little is known about higher \emph{multiplicity} amplitudes in strong backgrounds, even in the archetype example of plane wave backgrounds: the non-trivial spacetime dependence of the background massively increases the difficulty of even tree-level calculations. Higher multiplicity is important for QED cascades \cite{Mironov:2021fft,Mercuri-Baron:2024rdc}, but calculations 
are mostly approximated using the `incoherent product' assumption which puts all internal lines on-shell mentioned in Sec. 1.

\textbf{ 3.2 B) Higher loops and the Ritus-Narohznyi Conjecture:} in the Furry picture, loop corrections correspond to higher powers of $e$ (beyond leading order), but with $\xi$ treated exactly at each perturbative order. It has however been conjectured, based on the `constant crossed field' (CCF) model of a laser pulse, that extreme intensities can also enhance the dynamical coupling such that, effectively, each loops adds a factor of not $\alpha$, but $\alpha \chi^{2/3}$. This `Ritus-Narozhny conjecture'~\cite{Ritus1,Narozhnyi:1980dc,Fedotov:2016afw} implies, at least in a CCF, that the Furry expansion breaks down meaning an all-order resummation must be performed. As such, there are no theory predictions in the literature for the behaviour of physical quantities in the very high intensity regime -- it is simply not known how QED behaves there. 

It has been confirmed that `bubble chain' corrections $\mathcal{M}_n$ to electron forward scattering indeed scale asymptotically order-by-order with the number of loops $n$ as $\mathcal{M}_{n+1}/\mathcal{M}_n \sim \alpha\chi^{2/3}$~\cite{Mironov:2021ohk}. Intriguingly though, the resummed bubble chain remains unbounded as $\chi$ grows, hence (by unitarity) these cannot be the only relevant loop corrections at large $\chi$.  Explicit examples suggest that an analogous breakdown of perturbation theory can be found in fields other than a CCF, and \emph{even in classical physics}, though quantities scale with a different power of intensity~\cite{Heinzl:2021mji}. Other examples, however, show that there are also fully quantum cases where the Furry expansion remains convergent and well-approximated by its lowest order terms~\cite{Ekman:2020vsc}. In a constant magnetic field, it is possible to determine the strong field limit of the Heisenberg-Euler Lagrangian to all loops; these corrections exhibit a different field-depdence to those seen in CCF calcualations. Interestingly, the all-loop result is fully determined by one-particle \emph{reducible} contributions~\cite{Karbstein:2019wmj}. 

These disparate results tell us we must take a broader perspective on the Ritus-Narozhny conjecture -- as well as pushing CCF calculations to higher precision, we should look for analogous behaviour in other backgrounds, in other theories and regimes where the CCF model (or its extension in the locally constant field approximation (LCFA)) does not apply, and so on. This is a very challenging topic, and significant progress will likely require novel insights or the development of new theory tools. 

\textbf{ 3.2 C) Resummation:} connected to the above, we seek methods by which to re-sum a variety of corrections to SFQED processes. The resummation of leading IR logs and cancellation of IR divergences has been understood~\cite{Ilderton:2012qe}, but there are relatively few results on the resummation of finite IR terms, collinear degeneracies~\cite{Edwards:2020npu}, or loop corrections~\cite{Karbstein:2019wmj,Mironov:2021ohk}.  
For backgrounds of long duration, a `gluing approach' is available to build high-multiplicity and high-loop diagrams from a minimal set of leading contributions, valid even at moderate intensity where the LCFA fails~\cite{Dinu:2019pau}. This has been applied to processes such as nonlinear trident~\cite{Torgrimsson:2022ndq}, and has generated several new results, in particular an \emph{all-orders} (all loops and all multiplicity) derivation of classical radiation reaction directly from QED~\cite{Torgrimsson:2021wcj}.

Physics can benefit greatly from the import of new mathematical ideas and methods. Resurgence and trans-series are particularly interesting for quantum field theories. An important concept here is that working with only perturbative information, resurgent methods allow for the reconstruction of \emph{non-perturbative} information. Precision tests of this idea have been made for strong, even inhomogeneous, fields, in the context of the Euler-Heisenberg action~\cite{Dunne:2021acr, Dunne:2022esi}.  Trans-series structures can also be seen explicitly in e.g.~spin contributions to quantum radiation reaction~\cite{Torgrimsson:2024xyo}.

\textbf{ 3.2 D) Beyond background fields and plane waves:} almost all theory calculations in SFQED, and by extension theory input into Particle-In-Cell (PIC) simulations, is based on modelling the laser as a background plane wave (or even its zero frequency limit, the constant crossed field). Accounting analytically for focussing and going beyond the background field approximation are two of the most challenging theory problems in the field.

Focussing (i.e.~spatial field inhomogeneity) opens the door to richer physics~\cite{Fedotov2009exact,Gies:2013yxa,Gonoskov:2013ada,DiPiazza:2013vra,Heinzl:2017zsr,Karbstein:2019dxo,DiPiazza:2020wxp,Heinzl:2024cia}, but calculations based on the Furry expansion fail at, essentially, the first hurdle, because the Dirac equation cannot be solved for a realistic, intense, focussed pulse. To address this one can consider, broadly, two approaches. First, identify other approximations with which to construct particle wavefunctions, without using perturbation theory in $\xi$. Second, identify (special) cases in which one can find, by construction or by accident, exact solutions. Both approaches have yielded progress, the former in the context of high-energy approximations~\cite{DiPiazza:2013vra}, the latter using techniques from integrable systems~\cite{Heinzl:2017zsr}. 

Going beyond the background field approximation presents a conceptual challenge, as to date almost all SFQED calculations have been performed \emph{within} this framework (for exceptions based on lattice approaches see e.g.~\cite{Hebenstreit:2013qxa,Kasper:2015cca}).  In the background field approximation the laser is unaffected during interactions; it acts as an (infinite) reservoir of available energy which contributes to (or in the case of e.g.~nonlinear Compton \emph{stimulates})  processes, but evolves in time according only to Maxwell's equations in vacuum. While this sounds unphysical, there is a concrete re-interpretation of scattering in backgrounds in terms of scattering of coherent states of light~\cite{Frantz:1965,Kibble:1965zza}, hence background-field amplitudes are sensible quantities. Furthermore, the actual level of, e.g.,~beam depletion in laser experiments is estimated to be irrelevant due to the very large flux of photons in the beam, and the low density of probe electron beams~\cite{Seipt:2016fyu}. Studies are nevertheless motivated by future relevance in extreme regimes, and the Ritus-Narohznyi conjecture.

\vspace{-2mm}
\addcontentsline{toc}{section}{3.3}
\subsection{Simulation Objectives}
\vspace{-2mm}

As experiments make the transition from ``discovery'' to ``precision'', it will be necessary for simulations to include \emph{more accurate} models of QED processes, as well as \emph{additional} QED effects that, while not accessible today, will become visible with the next generation of experiments.
It is also essential to quantify the error that is made by simulations, in order to allow experiments to more closely  verify and constrain the properties of the underlying theory.

\textbf{3.3 A) Error determination:}
numerical simulations of SFQED processes are built upon a series of approximations, namely: the \emph{background field} approximation, a \emph{local} field approximation (either \emph{locally constant} or \emph{locally monochromatic}), and the \emph{cascade} approximation.
(In addition, where spin and polarisation are included as degrees of freedom, non-classical correlations between the spins of produced particles, present in the theory~\cite{Lotstedt:2009zz}, are neglected.)
These approximations arise from the underlying theory itself, and their validity is usually assessed by comparing the simulation result to a theoretical calculation, i.e. benchmarking~\cite{Harvey:2014qla,Blackburn:2018sfn}.
By definition, this kind of benchmarking can only be carried out in a parameter regime where it is actually possible to do the theoretical calculation.
As such, the error made by simulations is not known quantitatively where it is needed, specifically, in parameter regimes where only simulations can be used to make predictions for experiments.
A key objective in the development of numerical simulations is therefore to identify methods by which this error can be quantified.

\textbf{3.3 B) Better approximations:}
going beyond the current set of approximations, on which simulations are built, is fundamentally connected to the goal of placing error bounds on those simulations.
This is partially because comparing the `standard' and `improved' result when replacing an individual approximation can yield information about the true error made by the simulation framework.
As an example, the locally constant field approximation (LCFA), in which SFQED events are assumed to occur instantaneously, can be improved with post-local approaches that depend on field gradients~\cite{Ilderton:2018nws,DiPiazza:2018bfu} or assume that field is close to a monochromatic plane wave~\cite{Heinzl:2020ynb}.
Similar work is needed with regard to both the \emph{background field} and \emph{cascade} approximations.
Like the theory itself, simulations will assume that the total electromagnetic field may be separated into two components: a classical, coherent contribution at low frequency, which is unchanged by strong-field QED events, and an incoherent, quantised contribution at high frequency.
One objective is therefore to account for depletion of the background field~\cite{Seipt:2016fyu} and to allow for situations where the background and radiation fields are not well-separated in frequency.
Furthermore, simulations of high-order processes, like radiation reaction or EM showers, fundamentally rely on the cascade approximation, which posits that the dominant contribution to a high-order strong-field QED process is the incoherent product of its constituent first-order processes~\cite{Gonoskov:2021hwf,Fedotov:2022ely}.
Another objective is to find ways to model the `one-step' contributions (where multiple QED events occur within a single formation length) that are neglected under this approximation~\cite{King:2013osa}.

\textbf{3.3 C) Reproduction of experimental conditions:}
the rates and spectra of strong-field QED processes are highly sensitive to the spatiotemporal structure of the driving laser, as well as to the energy, emittance and spatial properties of any colliding electron beams.
These are often, if not always, modelled in idealised ways.
If a high-precision comparison to experimental data is to be made, it will not be sufficient to treat a focused laser pulse as an ideal paraxial Gaussian beam~\cite{beaurepaire.prx.2015} or the phase-space of the electron beam as uncorrelated~\cite{Magnusson:2023gbs}.
Many codes now make it possible to inject arbitrary EM pulses into the simulation domain~\cite{thiele.jcp.2016,viera.2024}, generally by determining the pulse's representation in Fourier space and propagating this to a given boundary.
The effect of complex laser structure on the expected results of a strong-field QED experiment has not yet been systematically studied.
Nevertheless, it is possible in principle for a simulation to match the laser pulse used in an experiment, provided complete information about the pulse's spatial and spectral properties is available.
Even without complete information, incorporating measurements of the laser transverse profile, duration and phase to improve the modelling of the background field (see, e.g. \cite{thevenet.2024}) in strong-field QED simulations is a key objective.

\textbf{3.3 D) New physics accessible at high precision or high quantum parameter:}
as the precision of experiments increases, it will become possible to investigate strong-field QED processes that are suppressed with respect to the cascade component, which is assumed to be dominant at high intensity and long pulse duration.
Apart from the `one-step' contributions discussed above in Sec. 3.2 A), e.g. one-step trident, this includes loop processes.
Some of these are already included in simulations: the one-loop mass operator, for instance, contributes to the evolution of the electron spin, which is modelled using a strong-field version of the T-BMT equation~\cite{li.prl.2020,Torgrimsson:2020gws}.
For consistency, it will be important for simulation codes to include other processes that occur at the same order in $\alpha$.
Moreover, as discussed in the \secref{sec:theoryObj}, these higher order contributions are conjectured to become as important as tree level in the extreme limit of $\alpha \chi^{2/3} \gtrsim 1$. A key objective is to solve the problem of simulating SFQED processes when the distinction between the cascade and non-cascade contributions is no longer meaningful. At the same time, simulations are already needed to a identify a pathway to achieve such a high value of $\chi$~\cite{yakimenko.prl.2019,Blackburn:2018tsn,Baumann:2018ovl}, and crucially what signals would provide evidence that it has been reached.

\vspace{-2mm}

\section{Readiness and expected challenges}
\vspace{-2mm}
A comprehensive and consistent experimental characterisation of the SFQED regime can only be achieved if all the collision parameters are known in detail. These include the spatiotemporal distribution of the laser electromagnetic fields with femtosecond and micron-scale precision and the phase-space distribution of the particle beam. 

High-quality and stable electron beams with ultra-relativistic energies can be provided by radio-frequency particle accelerators, such as the Stanford Linear Accelerator (SLAC) and the Eu.XFEL, able to accelerate electron beams with a maximum energy of 13 and 16.5 GeV, respectively. These accelerators are suited to drive x-ray Free Electron Lasers (FEL) and can thus provide superior beam qualities, including sub-percent energy spreads, micron-scale normalised emittances, and micron-scale longitudinal and transverse dimensions. 
FACET-II at SLAC is currently coupled with a relatively high-power laser system (peak power of the order of 10 TW) and a dedicated experimental campaign (E320 \cite{meuren2019probing}) is currently underway to study inverse Compton scattering and nonlinear BW in a moderately perturbative regime ($\xi \approx 1 -5$). A technical design report for comparagle experiments to be carried out at the Eu.XFEL has also been recently published by the LUXE (Laser Und XFEL Experiment) collaboration \cite{LUXE:2023crk,abramowicz2025inputesppuluxeexperiment}. LUXE is proposed as a permanent end-station at the Eu.XFEL, which will thus provide the unique opportunity to carry out long-term experimental campaigns specifically targeted at generating high-statistics datasets at high precision. The design and installation of a beamline to transport the electron beam to the LUXE experimental area have been funded by the European Union, but significant investment is still required for the installation and operation of a high-power laser and the necessary suite of detectors.   

Significant progress on laser wakefield acceleration of ultra-relativistic electron beams has also been reported, thus allowing in principle for SFQED experiments to be carried out in an "all-optical" setup at high-power laser facilities. For example, laser-wakefield accelerated electron beams with a maximum energy exceeding 10 GeV have been recently reported using a 500\,TW laser pulse \cite{picksley2024matched}. While the stability and spectral and spatial quality of wakefield accelerated electron beams can still not match those attainable with radiofrequency accelerators, proof-of-principle experiments demonstrate the possibility of finely controlling the beam parameters (see, e.g., Ref. \cite{Maier:2020}), thus opening up the possibility for high-precision studies of SFQED in high-power laser facilities as well. 

The quality and stability of the electron beam needs to be matched by those of the laser beam. Even small shot-shot fluctuations in the beam pointing, synchronisation, spatio-temporal distribution, and energy of the laser beam can result in non-negligible fluctuations in the intensity effectively experienced by the electron beam. The generally strong dependence of SFQED phenomena on field intensity can thus result in large fluctuations in expected outcomes, which thus significantly limits the possibility of carrying out detailed benchmarking of SFQED models against experimental results.  Plans to achieve \%-level monitoring and stability in the focussed laser intensity have been proposed (see, e.g., the Technical Design Report of the LUXE experiment \cite{LUXE:2023crk}), but require significant investment in laser technology to be translated into reality. Experimental programs in this area will be instrumental also for other applications of high-power lasers, such as plasma-based particle acceleration and generation of secondary photon sources.

Another issue that is common to both ``all-optical'' and ``hybrid'' (i.e. with electron beams from a radiofrequency accelerators) experiments is related to the detectors. These experiments require the simultaneous measurement of the spectral and spatial properties of particle and photon beams with a extremely wide range of energies and yield. For example, BW pair production requires single particle detection capability in a highly noisy environment, while Compton scattering would require detectors capable of measuring high-flux photon beams with a high photon energy. Several numerical and experimental studies on detector technology and background rejection have been recently reported (see, e.g., Ref. \cite{Salgado:2021fgt}). However, significant investment, dedicated access, and networking between different research communities are urgently needed to study SFQED with sufficient precision. 

Several smaller-scale laser facilities are also now operational across Europe in both research institutes and Universities. While these systems are generally not suited to directly access the SFQED regime, they do provide laser parameters suitable to carry out research and development in several technical aspects of a full-scale SFQED experiment, including stabilisation and high-precision characterisation of laser and particle beams and development of novel detectors. Moreover, facilities of this kind represent valuable hubs for the training of a pool of researchers, both at doctoral and postdoctoral level, that can then successfully undertake SFQED experiments at large-scale facilities.

\vspace{-2mm}
\section{Key recommendations and required support}
\vspace{-2mm}
Step-change progress in this emerging area of frontier fundamental physics can only be enabled by dedicated international programs devoted to addressing key technical and scientific challenges. Successful delivery of these programs is expected to produce tangible impact well beyond SFQED, in areas such as particle acceleration, generation and exploitation of novel particle and radiation sources, detector technology, and the design of the next generation of particle colliders. In the following, some key recommendations are outlined, together with the predicted level of support required.
\vspace{-2mm}
\addcontentsline{toc}{section}{5.1}
\subsection{Experimental Recommendations}
\vspace{-1mm}

\begin{enumerate}
\item \textbf{Installation and operation of experimental areas specifically devoted to SFQED and related high-intensity experiments}: while several international facilities can in principle provide field intensities sufficient to access SFQED regimes, specific investment is required to enable high-precision studies of SFQED in the laboratory. To do so, it is necessary to co-locate and synchronise different high-power laser beams within the same experimental area, and to enable co-location of high-power lasers with X-FEL radiation or high-energy electron beams from radio-frequency accelerators. As possible examples of this activity, hosting a high-power laser at the Eu.XFEL and coupling it with both the x-ray and electron beam will enable high-precision studies of vacuum polarisation (see, e.g., the HIBEF program \cite{HIBEF}), quantum radiation reaction, and BW pair production (see, e.g., the E320 program in the US \cite{meuren2019probing} and the proposal of the LUXE collaboration \cite{Abramowicz:2021zja,LUXE:2023crk,abramowicz2025inputesppuluxeexperiment}). Co-locating and synchronising the 1PW and 10 PW lasers at ELI - Beamlines would also allow for a unique capability worldwide to combine high-power and high-energy (1.5 kJ) laser beams. The scale and international interest for projects of this kind demands for large-scale investment at the European level.

\item \textbf{Dedicated access to high-power laser facilities for the stabilisation and high-precision characterisation of laser and particle beams}: high-power laser facilities generally award beam-time on a competitive basis, with particular attention being given to the ground-breaking and high-impact nature of the expected outcomes. Moreover, beam-time allocations rarely exceed 3-5 weeks. However, several technical advancements require longer periods of access, possibly by large and international collaborations. We would thus recommend designing beam access programs that are specifically targeting technical research and development, with associated expenses covered by transnational support. A similar program has been established for particle accelerators (see the ARIES project \cite{ARIES}) and could, for instance, be included as a specific access route within the Lasers4EU program \cite{Laser4eu}.

\item \textbf{Distributed investment in smaller-scale laser facilities and University laboratories for R\&D}: capillary development of small-scale laser laboratories at University level would strongly benefit the development of SFQED experimental research. Laser laboratories of this kind can operate as test-bed facilities for experimental techniques, novel detectors, and laser technology, while also providing the necessary training platform for the next generation of scientists. Besides their central role in SFQED, the radiation and particle sources delivered by facilities of this type can also be directly used for industrial and societal applications in areas as diverse as manufacturing, material characterisation, and healthcare. A coordinated plan across national research councils and specific calls for funding will enable developing and exploiting this type of facilities.

\item \textbf{Establishment of large-scale collaboration networks}: following the example set by the particle and accelerator physics communities (see e.g., the EuroNNAC \cite{EuroNNAC} and ALEGRO networks \cite{ALEGRO}), the interdisciplinary nature of this endeavor will strongly benefit from the formal establishment of international collaborations, with dedicated yearly workshops and funding for networking as well as the participation in the design of future particle accelerator facilities  \cite{Schroeder2023}. COST action funding and applications for dedicated international Doctoral Networks (e.g., via the Marie Skłodowska-Curie Actions) will enable establishing this type of networks.
\end{enumerate}
\vspace{-2mm}
\addcontentsline{toc}{section}{5.2}
\subsection{Theory Recommendations}
\vspace{-2mm}
Almost all theoretical calculations in SFQED - and, by extension, theory input into PIC simulations and experimental analysis - are based on modeling the laser as a background plane wave (or even a constant crossed field). New ideas and methods are needed if we are to make significant advances in regimes of high precision or ultra-high intensity. 

\begin{enumerate}
\item \textbf{Development of methods to efficiently calculate both higher multiplicity and higher-loop diagrams:}  beyond the CCF limit, a brute-force approach using standard models and methods seems unlikely to yield significant progress. Possible simplifications may be offered by going to the self-dual sector~\cite{Adamo:2025vzv}. Inverse methods may potentially have a role to play~\cite{Oertel:2015yma}, though to date there has been little exploration of their use in e.g.~nonlinear Compton scattering or nonlinear BW. Worldline, or first-quantised, approaches, certainly have a role to play, and can provide results which are not apparent or accessible in standard Feynman diagram approaches~\cite{DegliEsposti:2024rjw,Copinger:2024twl}.

 \item \textbf{Continued development of resummation methods:} this is not unrelated to the above and, in particular, we would need novel methods that allow for resummation of the Furry expansion at \emph{large} $\chi$. 
  In the CCF case it is possible to resum some sub-classes of diagram exactly, e.g. bubble chains~\cite{Mironov:2021ohk}, and this will likely be extended to other classes of diagrams eventually. It would be extremely useful to identify methods by which to identify which diagrams are \emph{relevant} at high $\chi$, ideally without computing all diagrams first. 

    \item \textbf{Development of an entirely different expansion scheme which holds in the high-$\chi$ regime}, or an approximation which can be trusted to effectively approximate the impact of higher-loop effects on a given process, or `blue-sky' goals. At present, it is not clear what form these would take, but they could be transformative for our understanding of QED in previously uncharted regimes.

  \item \textbf{Dedicated work for translating theoretical developments into simulation codes:} this might include implementation of more accurate methods, error estimation, verification and benchmarking.
    
\end{enumerate}
\vspace{-2mm}
\addcontentsline{toc}{section}{5.3}
\subsection{Simulation Recommendations}
\vspace{-1mm}

\begin{enumerate}
    \item
    \textbf{Establish a network to focus on error determination:}
    quantifying the theory uncertainty in SFQED simulations will require collaboration between researchers in theory, phenomenology and numerics.
    We recommend that a network of experts in these areas is established, with the aim of finding methods to quantify the uncertainty and how they should be implemented into simulation codes.

    \item
    \textbf{Create an open-source database of theory calculations and numerical methods:}
    identifying better approximations and verifying the performance of SFQED simulations requires benchmarking with theoretical predictions.
    We recommend that the numerical routines required to calculate these, as well as test case results, should be made publicly available, in the same way that many simulation codes are open-source.
    
    \item
    \textbf{Develop a start-to-end simulation framework for SFQED experiments:}
    we recommend starting developments towards an open-source, start-to-end simulation framework for SFQED experiments, which can import data from experimental measurements and beam-tracking software to model initial conditions, and includes synthetic detector responses.
    This could be integrated within existing frameworks used in high-energy particle physics~\cite{Ganis:2021vgv}.
\end{enumerate}

\vspace{-2mm}
\section{Conclusions}
\vspace{-2mm}
State-of-the-art laser facilities are now able to generate laser pulses with peak power approaching 10 PW and focussed intensities up to $10^{23}$ Wcm$^{-2}$. Recent pilot experiments are starting to demonstrate the potential of these facilities to access quantum electrodynamics in the strong-field regime in the laboratory. This research area is now sufficiently mature to transition from qualitative observational experiments to quantitative and high-statistics precision measurements. This is of paramount importance not only for several areas of fundamental physics but also for the optimisation of unique particle and radiation sources that these laser systems promise to deliver and for design of future multi-TeV lepton colliders. However, the full potential of multi-PW laser facilities can only be realised with significant investment in several areas of experimental, computational, and theoretical physics. This document sets out the main objectives and recommendations from the research community to carry out the first systematic and high-statistics studies of strong-field quantum electrodynamics in the laboratory.

\clearpage

\vspace{-2mm}
\section{Acknowledgments}
\vspace{-2mm}

G. Sarri acknowledges support from the EPSRC (Grant No. EP/V049186/1). B. King acknowledges support from The Leverhulme Trust, Grants No. RPG-2023-285 and No. RPG-2024-142.
A. Ilderton acknowledges STFC consolidator grant  ST/X000494/1  ``Particle Theory at the HiggsCentre''  and EPSRC grant EP/X024199/1, ``Fresh perspectives for QED in intense backgrounds: first quantised techniques in strong field QED''.
M. Wing acknowledges the support of DESY, Hamburg.
S.V.Bulanov wishes to acknowledge support by the NSF and Czech Science Foundation (NSF-GACR collaborative Grant No. 2206059 and NSF Grant No. 2108075 and Czech Science Foundation Grant No. 22-42963L.

This material is based upon work supported by the Department of Energy [National Nuclear Security Administration] University of Rochester ``National Inertial Confinement Fusion Program'' under Award Number(s) DE-NA0004144.
This report was prepared as an account of work sponsored by an agency of the United States Government. Neither the United States Government nor any agency thereof, nor any of their employees, makes any warranty, express or implied, or assumes any legal liability or responsibility for the accuracy, completeness, or usefulness of any information, apparatus, product, or process disclosed, or represents that its use would not infringe privately owned rights. Reference herein to any specific commercial product, process, or service by trade name, trademark, manufacturer, or otherwise does not necessarily constitute or imply its endorsement, recommendation, or favoring by the United States Government or any agency thereof. The views and opinions of authors expressed herein do not necessarily state or reflect those of the United States Government or any agency thereof.

\newpage

\bibliography{bibliography}

\end{document}